\documentclass[conference]{IEEEtran}
\usepackage{cite}

\ifCLASSINFOpdf
   \usepackage[pdftex]{graphicx}
\else
   \usepackage[dvips]{graphicx}
\fi
\usepackage{bm}
\usepackage[cmex10]{amsmath}
\begin{document}
%
\title{A Virtual Queue Approach for Online Estimation of Loss Probability Based on MVA Theory}
%
%
%

\author{\IEEEauthorblockN{
Guoqiang~Hu, Yuming~Jiang, Anne~Nevin
}
\IEEEauthorblockA{
Centre for Quantifiable Quality of Service in Communication Systems\\
Norwegian University of Science and Technology\\
O.S. Bragstads plass 2E, N-7491, Trondheim, Norway\\
Email: \{guoqiang.hu, jiang, anne.nevin\}@q2s.ntnu.no}
}

\maketitle

\begin{abstract}
In network quality of service provisioning, premium services generally 
require to keep a very small loss probability, which is infeasible to measure 
directly. The proposed virtual queue scheme estimates the small packet 
loss probability of a real queueing system by measuring queue statistics in a 
set of separate virtual queues. A novel scaling property between the
real queue and the virtual queues is deduced on the basis of the maximum 
variance asymptotic (MVA) theory. The new scheme retains the high accuracy and 
wide applicability of the MVA method for aggregated traffic 
while avoiding the high computational 
complexity in a direct application of the original MVA analysis in real 
time. This makes it suitable for online measurement applications such as 
network performance monitoring and measurement-based admission control.  
\end{abstract}


%
\IEEEpeerreviewmaketitle

\section{Introduction}
In the framework of quality-of-service (QoS) provisioning in the Internet, 
online QoS estimation (or monitoring) plays a significant role in 
dynamic resource allocation and measurement-based admission control.
As one of the most important QoS parameters, the packet loss probability 
(or buffer overflow probability) should have very small values in practical 
network operations to provide satisfying services, e.g., 
in the magnitude of $10^{-6} \sim 10^{-5}$. However,
a direct online measurement of such a small loss probability is difficult 
due to the large variability of the measurement values. 

QoS estimator based on virtual queue (VQ) 
\cite{Hu-Courcoubertis1995IEEEComm,Hu-Mao2005ComComm}
treats loss probability as a function of the buffer size 
when the input traffic remains the same. A specific type of function is presumed
for the loss probability. Then, separate virtual queues with reduced 
buffer size are derived from the real queue. A copy of the real input traffic
is injected to each of the virtual queues. As a result, a virtual queue sees
a relatively large overflow probability that is feasible to be measured. 
On the basis of the measured loss probabilities in virtual queues,
the loss probability of the real system can be calculated by interpolation 
according to the presumed shape of the loss probability function.
VQs are implemented by means of counters/timers, meaning 
low realization overhead. However, the queue responses in the Internet
are complex and dynamic. This makes the assumption of knowing a priori 
the loss probability pattern rigid or infeasible. As a consequence,
the applicability of such VQ schemes is much constrained.

A more general estimation approach is to measure
the statistics of input traffic, from which the packet loss probability 
is further computed by queueing analysis.
Generally, the more accurate the estimation, the more traffic 
statistics are necessary and the higher complexity the computation. 
Maximum variance asymptotic (MVA) \cite{Hu-Choe1998, Hu-Kim1998ACM} analysis 
provides queueing performance estimation when the input 
traffic can be modeled by a Gaussian process. This is typical in high-speed
links where the traffic is aggregated from a large number of flows.
The high accuracy of the MVA analysis was verified \cite{Hu-Choe1998, 
Hu-Kim1998ACM, Hu-Knightly1999IEEENetwork} for a wide spectrum of traffic
types including the long-range-dependent (LRD) traffic \cite{Hu-Willinger2002}. 
This admirable feature is attributed to the generality of the Gaussian
process and the central limit theory. However, the MVA analysis requires to 
measure the variance of traffic arrivals on different time scales. Among these, 
the most relevant time scale needs to be identified by search operation
with a complexity of $O(n)$, where $n$ is the number of discrete time scales.
The implementation complexity for the measurement and computation is
high in real-time application environments.

In this paper, we propose a VQ estimation scheme. While the proposed scheme
also has a low implementation complexity as general VQ approaches
\cite{Hu-Courcoubertis1995IEEEComm,Hu-Mao2005ComComm}, it makes no assumption 
on knowing a priori the changing pattern of the loss probability with buffer 
size. Instead, a novel scaling property is adopted, 
which is derived on the basis 
of the MVA analysis to map between the VQs and the real queue. Particularly,
the configuration of VQs and the mapping between loss probabilities 
in real/virtual queues are both guided by the scaling property.
The scheme thus inherits the high accuracy and wide applicability of the MVA 
method for aggregated traffic while avoiding the complexity 
in a direct application of the original 
MVA analysis in real time. Heuristic analysis shows that the proposed 
scheme achieves a significant variance-reduction-effect in estimating very 
small loss probability.

In the rest of this paper, Section II reviews the original MVA analysis 
of buffer overflow probability.  
The new VQ estimator is introduced and analyzed in Section III. 
Section IV evaluates the performance of the VQ estimator by simulation. 
The paper is concluded in Section V.

\section{Maximum Variance Asymptotic Analysis}

Consider a single server system with a first-in-first-out (FIFO) buffer 
of infinite size. The service rate is $c$. Let $Q$ denote the 
queue length. $P\{Q>q\}$ is the tail probability of the queue length. Here,
the parameter $q$ will be referred to as the \emph{threshold}.

Let $A(t)$ denote the amount of traffic arrival within an arbitrary 
time interval of length $t$. The mean traffic rate is $r$. 
The time-scale-dependent variance of the traffic is further defined as 
$v(t)=\text{VAR}[A(t)]$. In case the traffic is aggregated from a large number 
of sources, according to the central limit theorem the distribution of 
$A(t)$ can be approximated by a Gaussian distribution. In light of this, 
the tail probability 
$P\{Q>q\}$ is dominantly influenced by the variance $v(t)$ on a 
specific time scale $\tau$ called 
the dominant time scale (DTS) \cite{Hu-Choe1998}:
\begin{eqnarray}
\label{eq_relevant_TS}
\tau = \arg\inf_{t \ge 0} g(q,c,t) \text{ where } \\\nonumber
g(q,c,t)= \frac{q + (c-r)\cdot t}{\sqrt{v(t)}}.
\end{eqnarray}
In other words, $\tau$ is the value of $t$ that minimizes the function
$g(q,c,t)$. It can be further obtained by the MVA analysis \cite{Hu-Choe1998}:
\begin{eqnarray}
P\{Q > q\} &\approx& \exp(-\inf_{t \ge 0}\frac{g(q,c,t)^2}{2})
\label{eq_mva_delay} \\
&=&\exp(-\frac{g(q,c, \tau)^2}{2}).\nonumber
\end{eqnarray}

While Eq.~(\ref{eq_mva_delay}) provides a tight approximation of the tail 
probability $P\{Q>q\}$ for infinite buffer size, in practice the loss 
probability (or buffer overflow probability) is mostly concerned because the 
real buffer size is always bounded to limit the realization cost or to 
restrict the maximal queueing delay. Let $P_\text{L}(x)$ denote the
loss probability in a buffer of limited size $x$. In \cite{Hu-Kim1998ACM},
it is proposed to calculate $P_\text{L}(x)$ by: 
\begin{eqnarray}
P_\text{L}(x) &\approx& \frac{P_\text{L}(0)}{P\{Q>0\}} P\{Q>x\}.
\label{eq_mva_loss}
\end{eqnarray}
Here, $P\{Q > x\}$ and $P\{Q>0\}$ are interpreted in terms of tail probability
measures in the circumstance that the system would have infinite buffer size.
They are further calculated by Eq.~(\ref{eq_mva_delay}). $P_\text{L}(0)$ is the
loss probability if the system were bufferless.
$P_\text{L}(0)/P\{Q>0\}$ turns out to be a scalar. Its approximate solution is 
given in \cite{Hu-Kim1998ACM}, which relies on an integral operation. 

Despite the advantages in the accuracy and generality, 
the MVA analysis with Eq.~(\ref{eq_relevant_TS}-\ref{eq_mva_loss}) 
involves measurements of the variance process $v(t)$ over an 
extensive range of time scale $t$ 
\cite{Hu-Eun2003TON} as well as the search for DTS to calculate the infimum.
In practice, $v(t)$ can only be measured on discrete time scales
$\Delta, 2\Delta,...,n\Delta$ where $\Delta$ is the smallest
time scale that is relevant to the queueing performance.
Typically, $\Delta$ is in the magnitude of single packet service time,
which is very small on high speed links. This results in a large $n$.
The computational complexity in updating the statistics $v(t)$ amounts
to $O(n)$. So does the complexity in search of the DTS. 
Consider the transmission of video transport streams with
standard packet size of 188 bytes over a 100~Mbps link. 
$n$ amounts to 6649 in order to cover time scales up to 100~ms. 
This is very demanding in the implementation for online estimation.

\section{Virtual Queue Estimator}

The proposed VQ estimator aims to determine the very small loss probability
$P_\text{L}(x)$ for a single server system with a FIFO buffer of size $x$. 
This is called the \emph{real system} in the following sections.
MVA analysis is applied to derive a novel scaling property. By taking advantage 
of the scaling property, the loss probability can be computed from
three probability statistics respectively measured in three VQs.
Consequently, the measurement of the variance process $v(t)$ and 
the search for the DTS are not necessary any more.

In this section, the basic concept of the VQ estimator is firstly addressed.
The scaling property is derived by MVA analysis.
Then, a heuristic analysis is performed to show the variance reduction
achieved by the VQ estimator. On this basis, a guideline for the parameter
configuration is outlined.

\subsection{VQ Concept and Scaling Property}

To apply Eq.~(\ref{eq_mva_loss}) in the VQ paradigm, a total of three VQs are
constructed, denoted by VQ1, VQ2 and VQ3. They are configured to obtain the 
measure $P_\text{L}(0)$, $P\{Q>0\}$ and $P\{Q > x\}$, respectively.
The input traffic in each VQ is a copy of the real input traffic. Among the
three VQs, VQ1 and VQ2 have the same service rate $c$ 
as the real system. However,
VQ1 is bufferless so that the loss probability $P_\text{L}(0)$ can be easily
measured. VQ2 has an infinite buffer size in order to give the measurement 
of $P\{Q>0\}$.
Note that $P_\text{L}(0)$ and $P\{Q>0\}$ have relatively large values. So,
the correspondent measurements in the VQ1 and VQ2 are straightforward.
As a result, the scalar $P_\text{L}(0)/P\{Q>0\}$ in Eq.~(\ref{eq_mva_loss})
is obtained from VQ-measurements. The complex numerical calculation 
in \cite{Hu-Kim1998ACM} is omitted.

VQ3 has an unlimited FIFO buffer for the estimation of the tail 
probability $P\{Q > x\}$, which is too small to be measured directly.
The idea is to scale down the threshold $x$ and service rate $c$ to 
build the VQ3. The resulting tail probability in VQ3 is thus
raised to a level that can be practically measured. In the following, 
we use $x'$, $c'$ and $P\{Q'>x'\}$ to denote the concerned threshold, 
service rate and tail probability in VQ3, respectively. 
The key to the design lies in mapping
the measured tail probability $P\{Q'>x'\}$ in VQ3 to the tail
probability $P\{Q > x\}$ according to a scaling property.

To avail a scaling between $P\{Q > x\}$ and  $P\{Q'>x'\}$ without the traffic 
measurement for $v(t)$ and the calculation of DTS, it is necessary to keep 
an unchanged DTS in VQ3 after the scale-down. With reference to
Eq.~(\ref{eq_relevant_TS}), we found that this can be assured if
\begin{eqnarray}
g(x',c',t)=k~g(x,c,t),
\label{eq_condition}
\end{eqnarray}
where $k$ is a scalar independent of $t$. This condition is fully satisfied
if we set:
\begin{eqnarray}
\label{eq_vq_qlength}
x'&=&\frac{x}{\alpha}, \\
\label{eq_vq_channel}
c'&=&\frac{c}{\alpha}+(1-\frac{1}{\alpha})r,
\end{eqnarray} 
where $\alpha:\alpha>1$ is the constant \emph{scaling factor} and $r$ 
is the mean traffic rate. It leads to
$g(x',c',t)=g(x,c,t)/\alpha$. Inserting this into Eq.~(\ref{eq_mva_delay}),
we obtain the scaling property between $P\{Q > x\}$ and  $P\{Q'>x'\}$
\begin{eqnarray}
P\{Q'>x'\} &\approx& \exp (-\frac{1}{\alpha^2}\inf_{t \ge 0}\frac{g(x,c,t)^2}{2}) \nonumber\\
&\approx& P\{Q>x\}^{1/\alpha^2}.
\label{eq:dev_mapping}
\end{eqnarray}
The inverse of Eq.~(\ref{eq:dev_mapping}) leads to
\begin{eqnarray}
P\{Q>x\} \approx P\{Q'>x'\}^{\alpha^2}.
\label{eq:mapping}
\end{eqnarray}

Eq.~(\ref{eq:mapping}) reveals how the tail probability $P\{Q > x\}$ is 
mapped from the measurement of $P\{Q' > x'\}$ in VQ3. Together with the
measurements $P_\text{L}(0)$ and $P\{Q>0\}$ from VQ1/VQ2, the loss
probability $P_\text{L}(x)$ of the real system is calculated according to  
Eq.~(\ref{eq_mva_loss}). The overall computational complexity is as low as
$O(1)$ in contrast to that of the MVA analysis. 

\subsection{Evaluation of Variance Reduction}

Because $P_\text{L}(0)$ and $P\{Q>0\}$ in Eq.~(\ref{eq_mva_loss}) can be
measured in VQ1/VQ2 with relatively high credibility as described 
in Section III.A, 
the variance in the estimation of $P_\text{L}(x)$ is mainly 
attributed to the estimation of the very small tail probability 
$P\{Q > x\}$. In the following analysis, we focus on the effect of 
$P\{Q > x\}$, neglecting that of $P_\text{L}(0)$ and $P\{Q>0\}$.
The variance reduction of the VQ estimator is thus 
characterized by the ratio $\eta$:
\begin{eqnarray}
\eta = (\frac{\text{COV}[\hat{P}^{\text{VQ}}\{Q>x\}]}{\text{COV}[\hat{p}_x]})^2.
\label{eq_ratio1}
\end{eqnarray}
Here COV[$\cdot$] denotes the coefficient of variation (COV). 
$\hat{p}_x$ stands for the direct measurement of $P\{Q > x\}$. 
$\hat{P}^{\text{VQ}}\{Q>x\}$ denotes the estimation of 
$P\{Q > x\}$ by the VQ approach. Because 
$E[\hat{P}^{\text{VQ}}\{Q>x\}] \approx E[\hat{p}_x]=P\{Q > x\}$,
Eq.~(\ref{eq_ratio1}) is simplified to
\begin{eqnarray}
\eta =\frac{\text{VAR}[\hat{P}^{\text{VQ}}\{Q>x\}]}{\text{VAR}[\hat{p}_x]},
\label{eq_ratio2}
\end{eqnarray}
where VAR[$\cdot$] denotes the variance.

Let $\hat{u}_{x'}$ denote the measurement of $P\{Q' > x'\}$ in VQ3.
According to Eq.~(\ref{eq:mapping}),
$\hat{P}^{\text{VQ}}\{Q>x\}= (\hat{u}_{x'})^{\alpha^2}$. Through linearization
\cite{Hu-Rice1995Book}, the variance of $\hat{P}^{\text{VQ}}\{Q>x\}$ is:
\begin{eqnarray}
&&\text{VAR}[\hat{P}^{\text{VQ}}\{Q>x\}] \nonumber \\
& \approx & \text{VAR}[\hat{u}_{x'}](\frac{d \hat{P}^{\text{VQ}}\{Q>x\}}{d\hat{u}_{x'}})^2|_{\hat{u}_{x'}=P\{Q'>x'\}} \nonumber \\
&=& \text{VAR}[\hat{u}_{x'}]\alpha^4 P\{Q'>x'\}^{2(\alpha^2-1)}
\label{eq_linear}
\end{eqnarray}

Formally, the tail probability $P\{Q > x\}$ is defined as the average
ratio between the sum of the time duration in which $Q>x$ holds and the total 
observation duration. For a heuristic analysis, however, we equalize this 
tail probability to the probability that an arbitrary packet sees $Q>x$ upon 
its arrival \footnote{The quantitative difference between this 
``packet-seen'' probability and the ``time-average'' probability is generally 
small.}. Correspondingly, we employ $\hat{p}_x=M/N$ where $M$ is the
number of packets seeing $Q>x$ and $N$ is a constant denoting the total number 
of packet arrivals. So, $E[\hat{p}_x]=P\{Q>x\}$.
Assume that the probability that a packet sees $Q>x$ is 
independent. $M$ then follows a binomial distribution. So, the variance of
$\hat{p}_x$ turns out to be:
\begin{eqnarray}
\text{VAR}[\hat{p}_x]=P\{Q>x\} (1-P\{Q>x\})/N.
\label{eq_variance1}
\end{eqnarray}
Similarly, the variance of $\hat{u}_{x'}$ in VQ3 is:
\begin{eqnarray}
\text{VAR}[\hat{u}_{x'}]=P\{Q'>x'\} (1-P\{Q'>x'\})/N.
\label{eq_variance2}
\end{eqnarray}

Applying Eq.~(\ref{eq:dev_mapping}) and inserting
Eq.~(\ref{eq_linear})-(\ref{eq_variance2}) to Eq.~(\ref{eq_ratio2}), we obtain:
\begin{eqnarray}
\label{eq_ratio_final1}
\eta &\approx& \alpha^4 \frac{P\{Q>x\}^{(1-1/\alpha^2)}-P\{Q>x\}}{1-P\{Q>x\}}.
\end{eqnarray}

\begin{figure*}[t]
  \noindent
  \begin{minipage}[b]{0.32\linewidth} 
    \centering
    \includegraphics[width=1\linewidth]{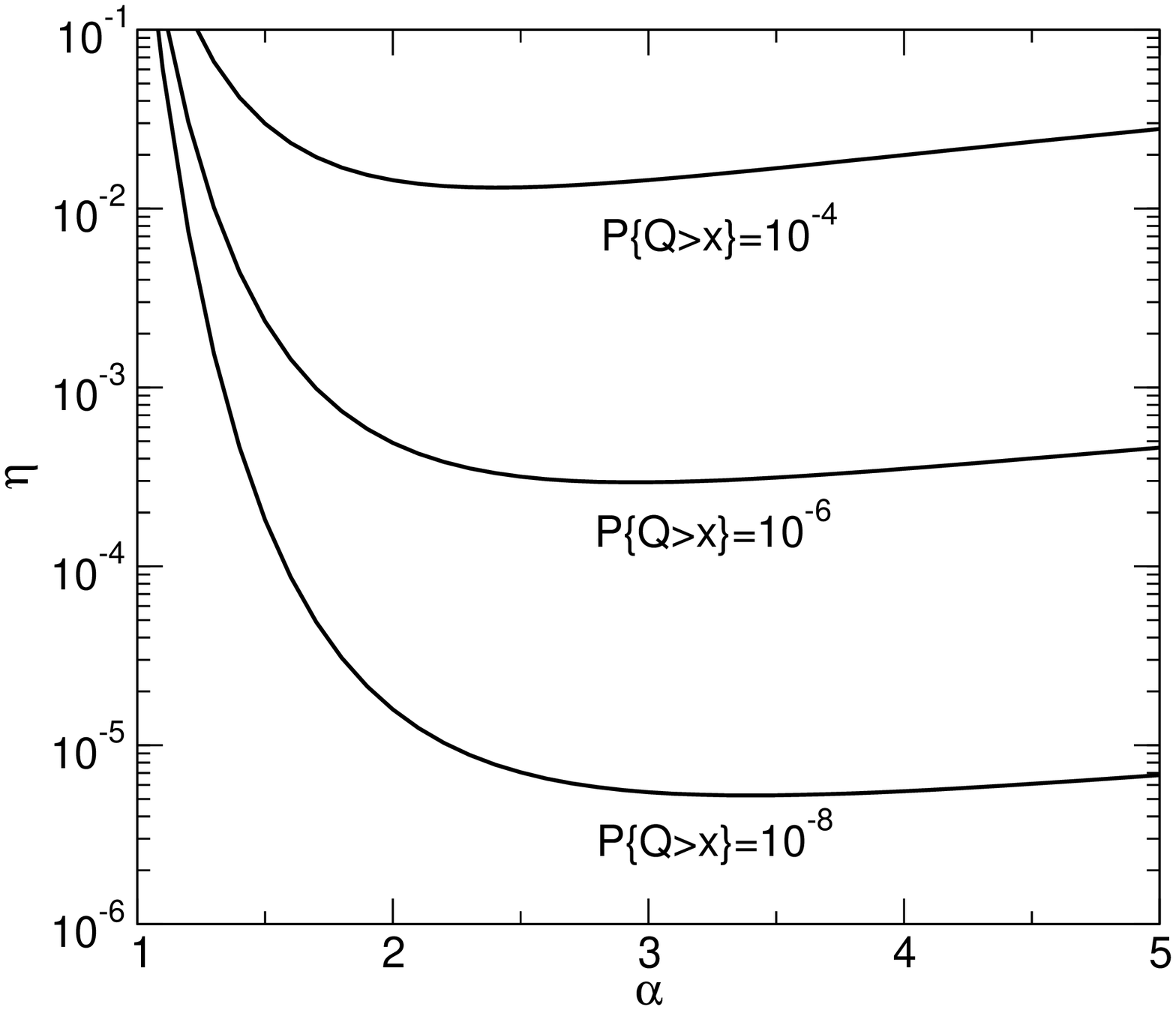} 
    \caption{Influence of $\alpha$ on the variance reduction.} 
    \label{fig_ratio} 
  \end{minipage}\hfill 
  \begin{minipage}[b]{0.32\linewidth} 
    \centering
    \includegraphics[width=1\linewidth]{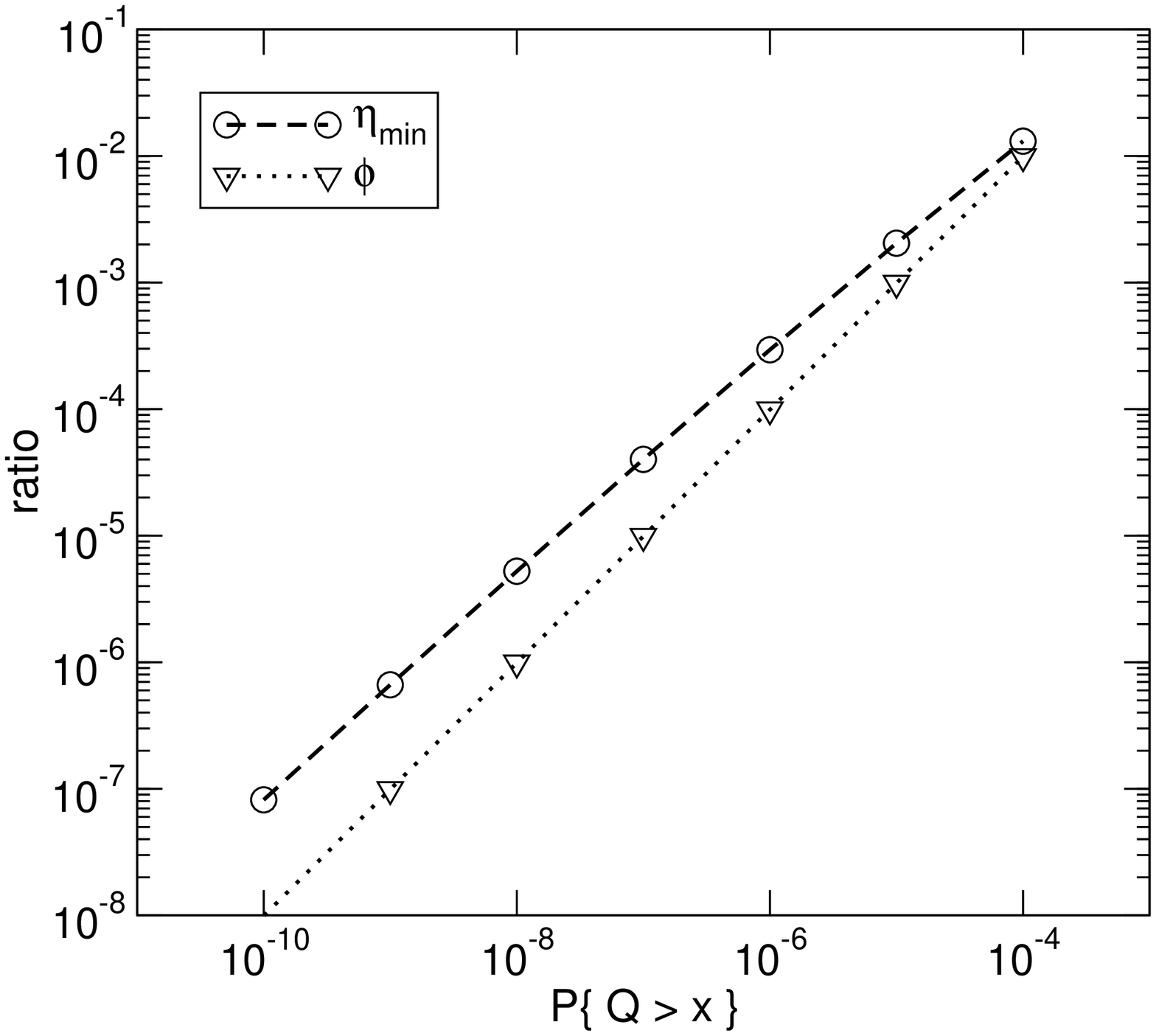} 
    \caption{The minimal $\eta$ at different levels of $P\{Q>x\}$.} 
    \label{fig_eta_min} 
  \end{minipage}\hfill 
  \begin{minipage}[b]{0.32\linewidth} 
    \centering
    \includegraphics[width=1\linewidth]{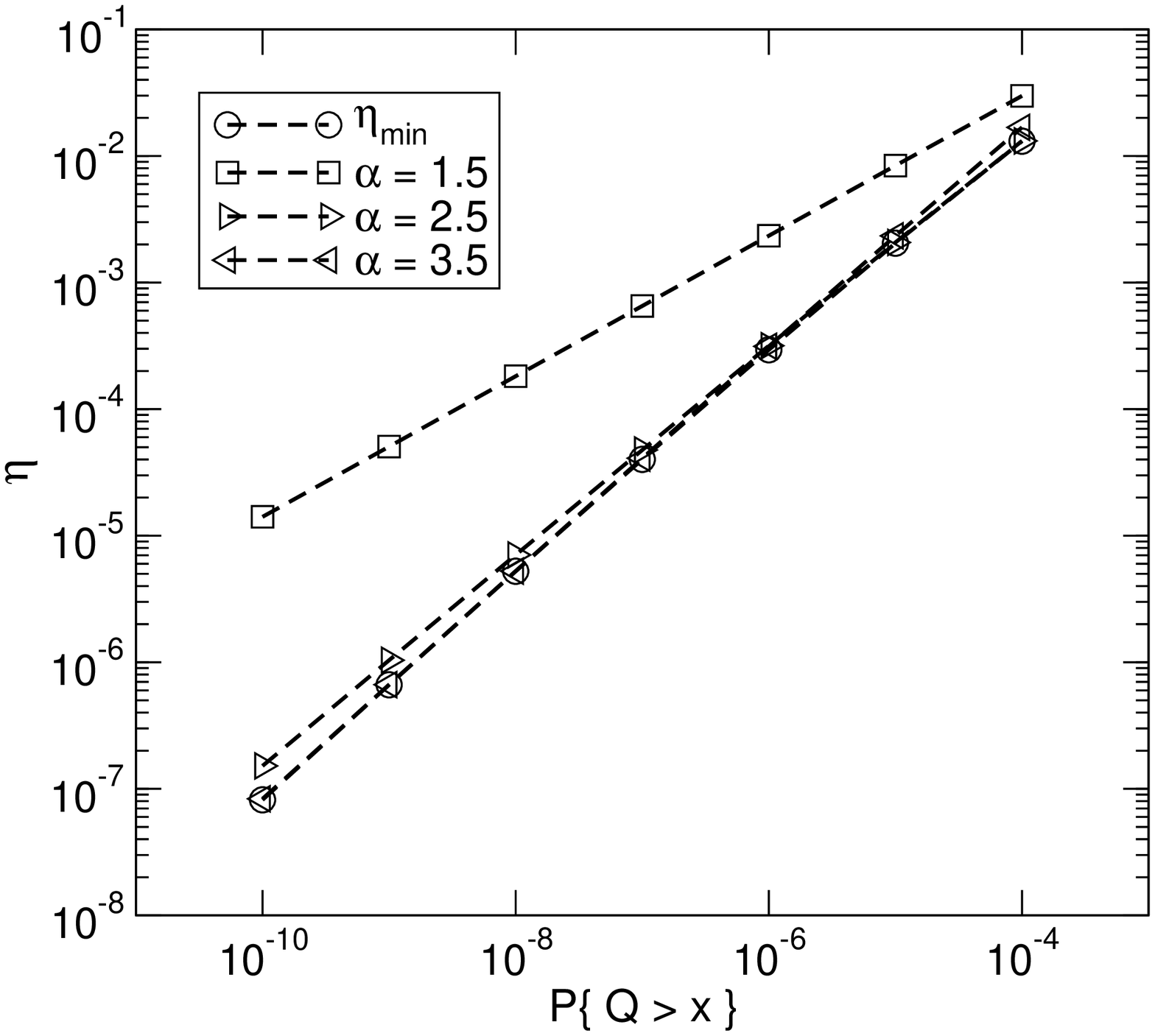} 
    \caption{$\eta$ with fixed values of $\alpha$.} 
    \label{fig_dev} 
  \end{minipage}\hfill 
\end{figure*}

In Fig.~\ref{fig_ratio}, $\eta$ is plotted with respect to the scaling factor
$\alpha$ according to Eq.~(\ref{eq_ratio_final1}). $P\{Q>x\}$ is set to
$10^{-4}, 10^{-6}, 10^{-8}$, respectively. 
At the beginning, $\eta$ decreases very fast with the increase of $\alpha$
because a large scaling factor results in a high level of $P\{Q'>x'\}$ which
allows for a stable measurement $\hat{u}_{x'}$. On the other hand, the mapping
operation Eq.~(\ref{eq:mapping}) can also amplify the variance of
$\hat{u}_{x'}$. Therefore, $\eta$ increases for too large values of $\alpha$.
However, such increase is very slow with the change of $\alpha$, meaning that
$\eta$ is insensitive to over-scaling. Actually for all three curves 
$\eta$ becomes quite stable as long as $\alpha \ge 2.0$.

With Eq.~(\ref{eq_ratio_final1}), the minimal $\eta$ can be numerically
calculated at different magnitudes of $P\{Q>x\}$, which is plotted in
Fig.~\ref{fig_eta_min} marked with legend $\eta_\text{min}$. It shows that 
the smaller the tail probability is, the more variance reduction can be 
achieved. This verifies the potential of the
VQ scheme for the estimation of rare events.

As a reference, we also look at a ratio 
$\phi=(\text{COV}[\hat{\xi}]/\text{COV}[\hat{p}_x])^2$. Here,  $\hat{\xi}$ 
is the direct measurement of a reference tail probability that has 
a relatively large value of 0.01. 
In Fig.~\ref{fig_eta_min}, the curve of $\phi$ locates quite close to the
curve of $\eta_\text{min}$. It implies that the stability of the VQ
estimation is comparable to the direct measurement of a probability in
the magnitude of 0.01, which is quite practical. This promises a good 
applicability of the VQ scheme in real-time environments.

\subsection{Configuration of the Scaling Factor}

With $\alpha\approx\sqrt{\log{P\{Q>x\}}/\log{P\{Q'>x'\}}}$ obtained from 
Eq.~(\ref{eq:mapping}), Eq.~(\ref{eq_ratio_final1}) is rewritten as:
\begin{eqnarray}
\label{eq_ratio_final2}
\eta \approx \frac{1-P\{Q'>x'\}}{P\{Q'>x'\} \log^2{P\{Q'>x'\}}}\cdot
\frac{P\{Q>x\} \log^2{P\{Q>x\}}}{1-P\{Q>x\}}.
\end{eqnarray}
Note that on the right hand side, the first term depends solely on 
$P\{Q'>x'\}$ and the second term depends solely on $P\{Q>x\}$.
Therefore, the minimization of $\eta$ with respect to $P\{Q'>x'\}$
needs only to consider the first term. By numerical computation,
it is found that $P\{Q'>x'\}=0.2032$ in order to achieve a minimal
$\eta$. On this basis, $\alpha$ can be determined correspondingly using
Eq.~(\ref{eq:mapping}) for a specific range of the tail probability 
 $P\{Q>x\}$.

In practice, the real $P\{Q>x\}$ is a dynamic parameter. A fixed setting 
of $\alpha$ should provide a stable estimation over a wide range of the
$P\{Q>x\}$ values. In Fig.~\ref{fig_dev}, we use Eq.~(\ref{eq_ratio_final1}) 
and show the deviation of the ratio
$\eta$ from the minimal $\eta$ when $\alpha$ is fixed to 1.5, 2.5 and 3.5,
respectively. It is seen that a large performance degradation only occurs
in case of too small $\alpha=1.5$. With sufficiently large $\alpha=2.5,3.5$,
$\eta$ is quite stable and very close to the optimum. This is in consistency
with the observations in Fig.~\ref{fig_ratio}.

\section{Performance Evaluation}

\begin{figure*}[htp]
  \noindent
  \begin{minipage}[b]{0.4\linewidth} 
    \centering
    \includegraphics[width=1\linewidth]{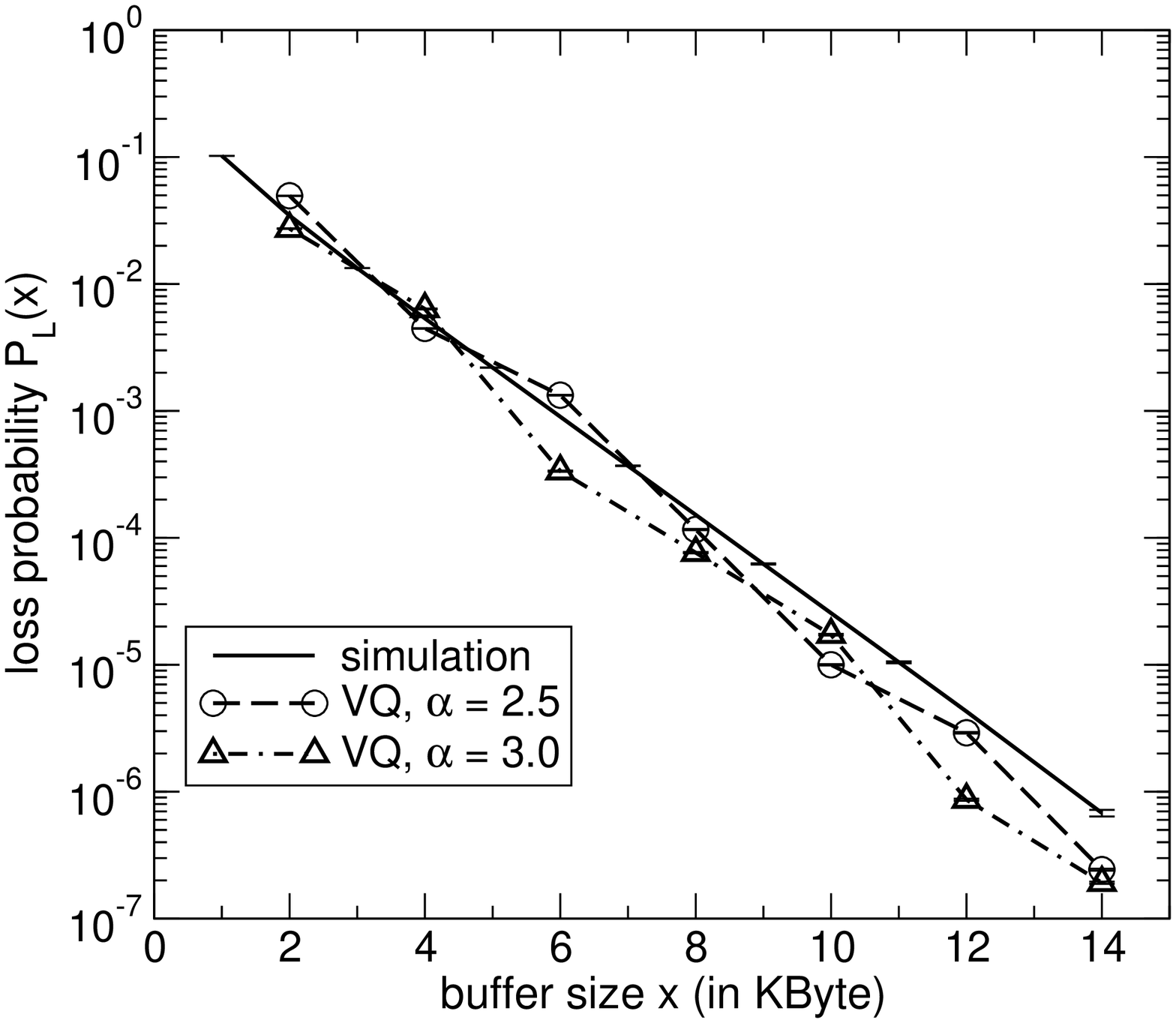} 
    \caption{Estimation accuracy for synthetic audio flows.} 
    \label{fig_ccdf_onoff} 
  \end{minipage}\hfill 
  \begin{minipage}[b]{0.4\linewidth} 
    \centering
    \includegraphics[width=1\linewidth]{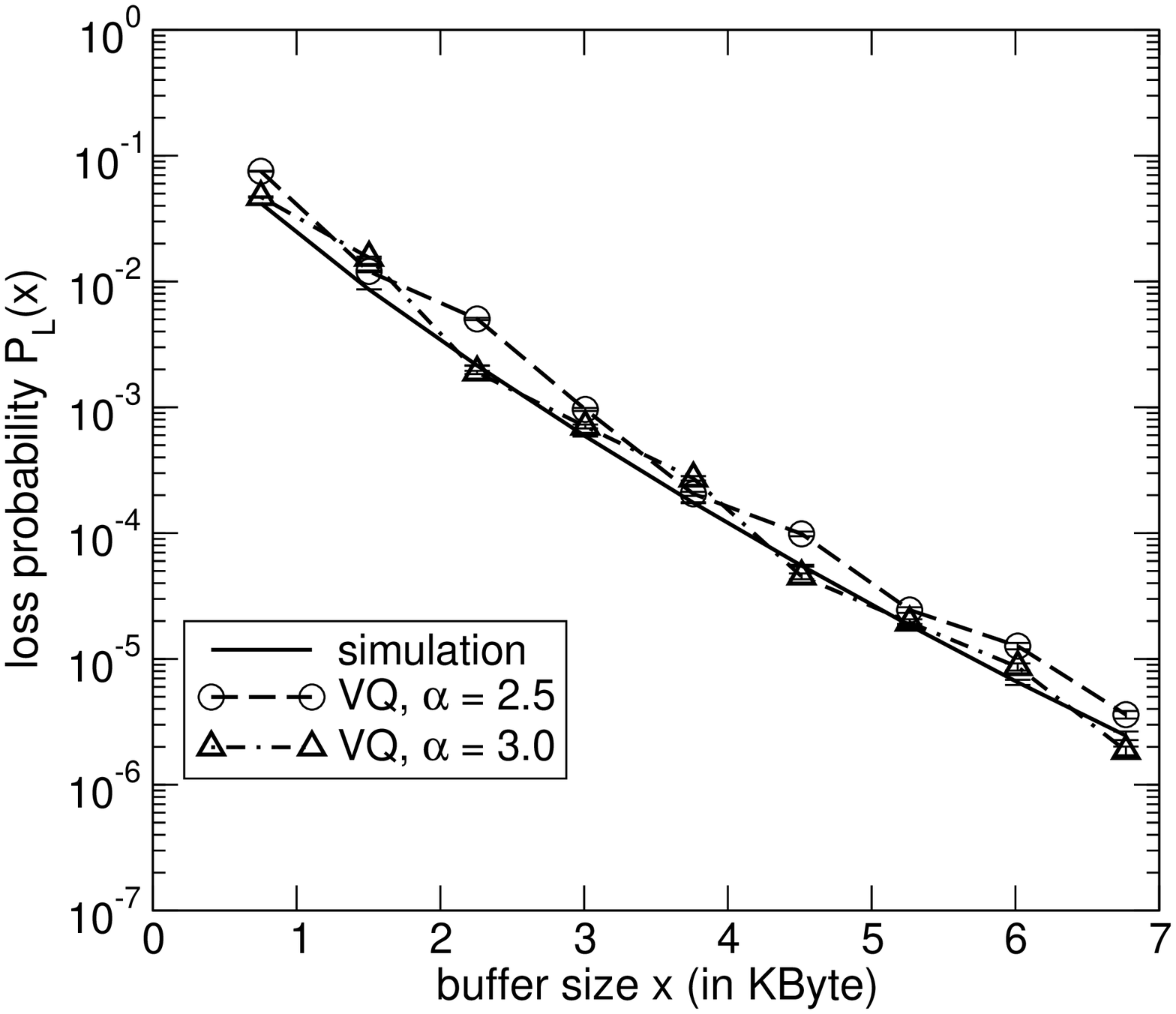} 
    \caption{Estimation accuracy for synthetic video flows.} 
    \label{fig_ccdf_mmpp} 
  \end{minipage}\hfill 
\end{figure*}

\begin{figure*}[htp]
  \noindent
  \begin{minipage}[b]{0.4\linewidth} 
    \centering
    \includegraphics[width=1\linewidth]{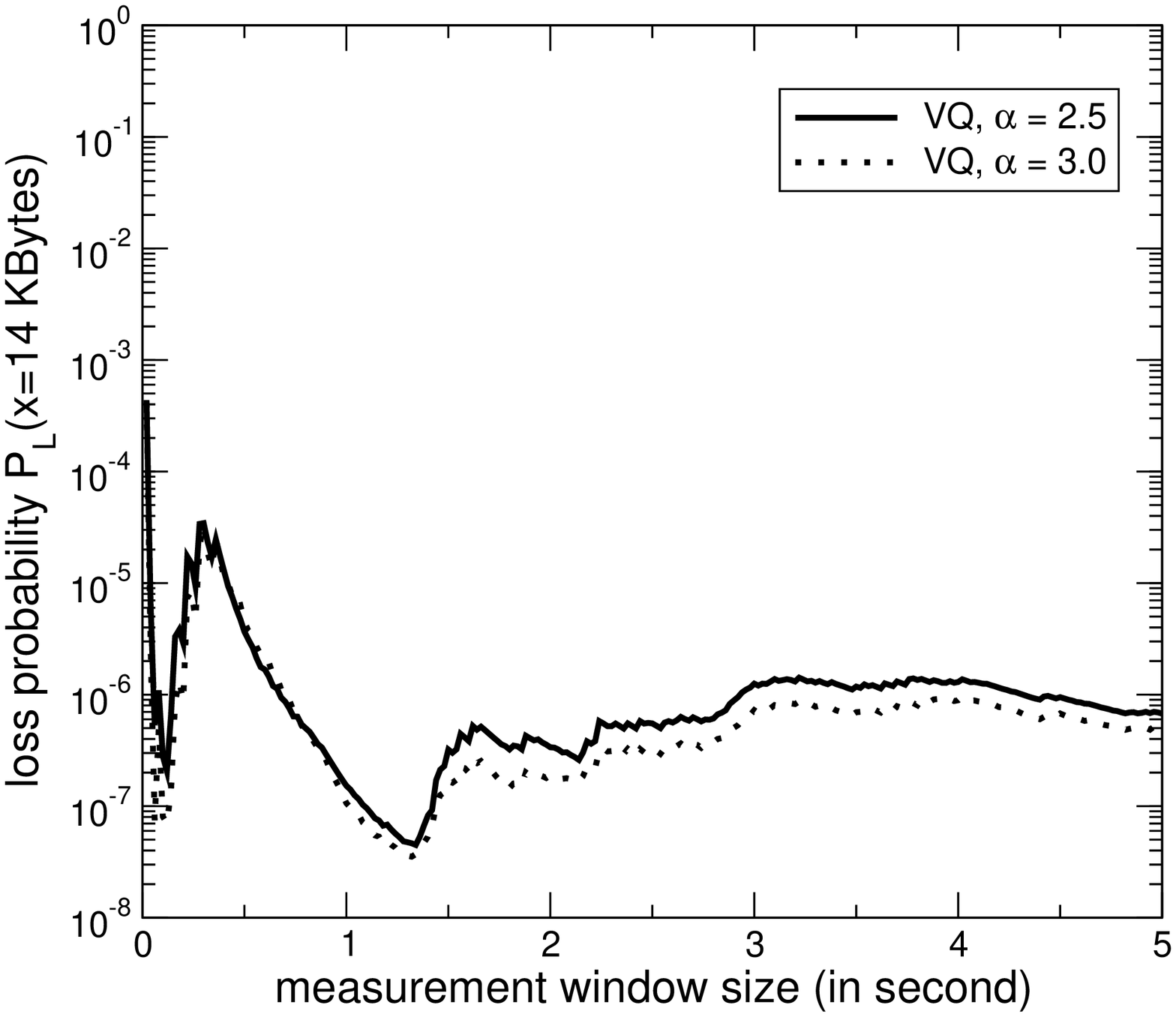} 
    \caption{Transient response for synthetic audio flows.} 
    \label{fig_transient_onoff} 
  \end{minipage}\hfill 
  \begin{minipage}[b]{0.4\linewidth} 
    \centering
    \includegraphics[width=1\linewidth]{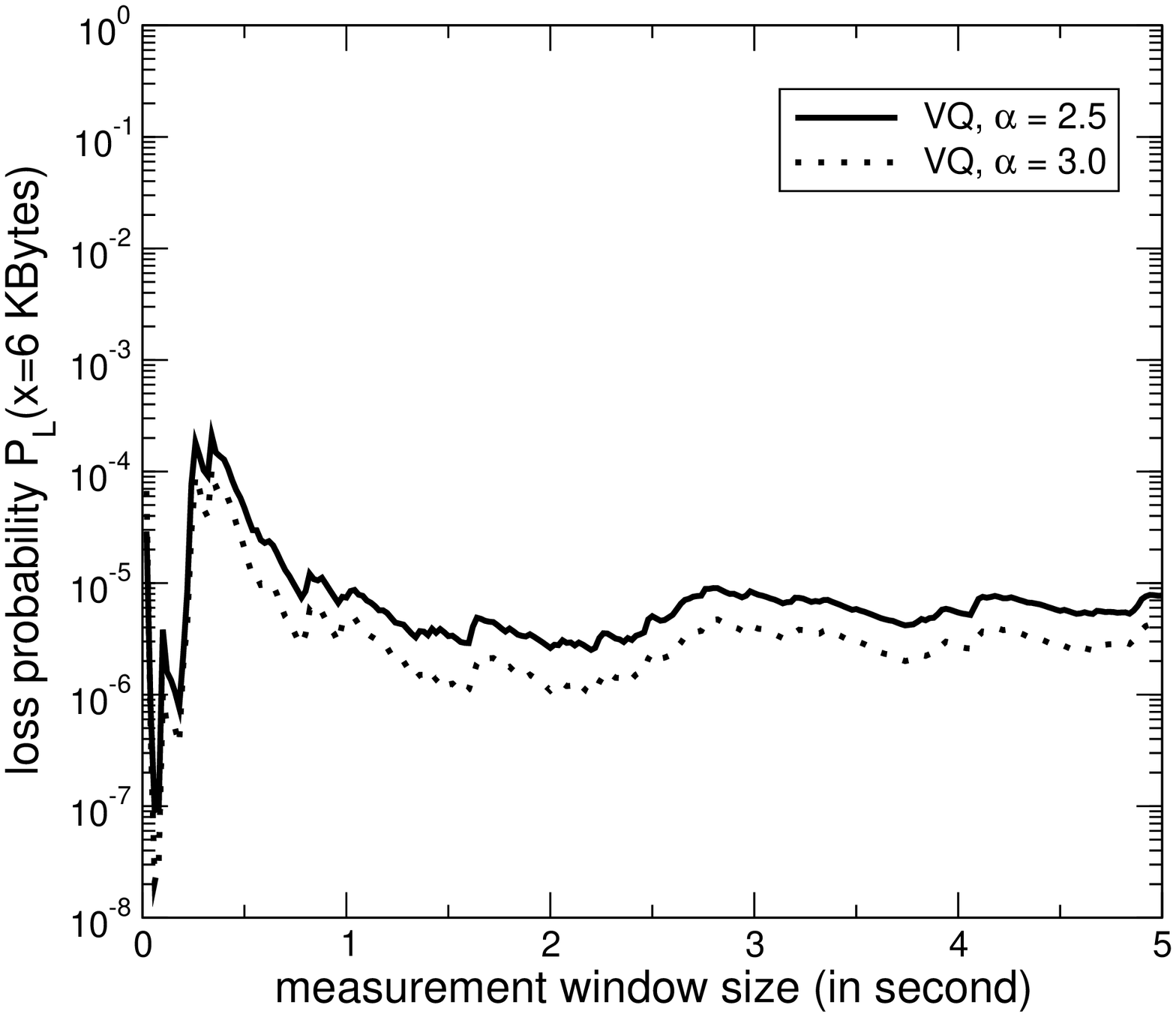} 
    \caption{Transient response for synthetic video flows.} 
    \label{fig_transient_mmpp} 
  \end{minipage}\hfill 
\end{figure*}

\subsection{Simulation Configuration}

The performance of the proposed VQ estimator is evaluated by simulation 
for a single-server FIFO queue system with bounded buffer size. 
The service rate is $c=100$ Mbps. The VQ estimator uses three VQs, 
in which the statistics $P_\text{L}(0)$, $P\{Q>0\}$ and $P\{Q'>x'\}$ are 
respectively measured for a common time duration. The duration is called 
\emph{measurement window}. Naturally, larger measurement window results in
more reliable measurement results. In VQ3, the buffer size $x'$ is configured 
by Eq.~(\ref{eq_vq_qlength})
with scaling factor $\alpha=2.5,3.0$ respectively.
The mean traffic rate $r$ is sampled in terms of the 
average rate over the elapsed time in the current measurement window. 
VQ3 uses samples of $r$ to periodically update its service rate $c'$ 
according to Eq.~(\ref{eq_vq_channel}) at constant interval of 1~ms. 
The real loss probability is estimated at the end of the measurement window by
Eq.~(\ref{eq_mva_loss}) and (\ref{eq:mapping}).
A measurement window always starts when the queue system is in the
steady state after a long transient simulation phase (5000~s).

Two types of input traffic are inspected. Type A traffic aims to simulate the 
traffic of audio services. It is the aggregation of homogeneous on-off 
sources. The on-period follows a negative exponential distribution with mean 
value 0.2~s. The off-period is negative-exponentially distributed with 
mean value 0.6~s. In the on-period, each source generates a bit stream at 
constant rate of 64~kbps. The bit stream is packetized into packets of constant 
length 500 bytes. 

Type B traffic is also a multiplexing of independent flows. Each flow is 
generated by a 3-states markov modulated Poisson process (MMPP) to simulate
the traffic of video services. 
We adopt the setting in \cite{Hu-Dan2006Tele}: the state-dependent rate 
$\lambda_1=83/s$, $\lambda_2=367/s$, $\lambda_3=661/s$; the state transition 
rate $q_{12}=40,q_{13}=1,q_{21}=10,q_{23}=1.3,q_{31}=6,q_{32}=0.1$. 
The packet length is constant 188 bytes. This setting was derived from realistic
MPEG-4 video traces with an average bitrate of 540~kbps \cite{Hu-Dan2006Tele}.
In either case (Type A/B), the system load $\rho$ is configured by changing the
number of traffic sources. 

\subsection{Estimation Accuracy}

The estimation performance is first evaluated by setting a very large
measurement window (25000~s) so that all the measurement results are stable. 
This aims to assess the estimation accuracy in steady states.
The loss probability estimations by the VQ estimator are plotted in
Fig.~\ref{fig_ccdf_onoff} and \ref{fig_ccdf_mmpp} for Type A 
(at load $\rho=0.8$) and Type B traffic (at load $\rho=0.81$), 
respectively. For comparison, the real loss probabilities measured by the long
simulations are drawn with solid lines.

In the figures, the VQ estimations show a good follow-up with the
real loss probabilities. The estimation deviation is well constrained within
one order of magnitude, which corresponds to the accuracy of the MVA method
\cite{Hu-Kim1998ACM}. This stands for a quite high accuracy among the existing 
estimation methods that are able to deal with generalized input traffic.

\subsection{Transient Response}

To illustrate the transient response of the VQ estimator, the loss probability 
under specific buffer size is estimated with the measurement window size 
constantly incremented at a step of 20~ms. The results are plotted for Type A
traffic in Fig.~\ref{fig_transient_onoff} and for Type B traffic in 
Fig.~\ref{fig_transient_mmpp}. The traffic setting is the same to that in 
Section IV.B.

In both graphs, the VQ estimator gives a quick response with probable
overshoot. The estimations fluctuate for a short time and converge soon.
Stable estimations are obtainable with a measurement window
as small as 3~s. This verifies the effectiveness of the VQ estimator for 
online loss probability estimation. The same evaluation is also 
performed for a direct measurement of the real loss probability. For the system
scenario corresponding to Fig.~\ref{fig_transient_onoff}, the direct 
measurement does not report any packet loss until 13~s. For the scenario
of Fig.~\ref{fig_transient_mmpp}, the direct measurement records the first
packet loss after 22~s. (The results are not shown in the graphs so as to focus
on the response time-scale of the VQ estimator). The performance advantage of 
the VQ scheme is substantial.

\section{Conclusion and Outlook}

The proposed VQ scheme does not rely on any presumption of the loss
probability pattern with respect to buffer size. Instead, a novel scaling
property is derived based on the MVA analysis to map
between the real queue and the VQs. The new scheme inherits the advantages of
conventional VQ approaches and the MVA method: low implementation overhead,
high accuracy and a general applicability for aggregated Internet traffic.
These advantages make it a promising online estimator.

Our work also opens a new way to connect the rich research field
of queueing analysis to the real-time network performance estimation. We are
looking forward to exploring further scaling properties to avail the
development of advanced VQ solutions.


%

\ifCLASSOPTIONcaptionsoff
  \newpage
\fi



%

\bibliographystyle{IEEEtran}

\bibliography{mbac}

\begin{thebibliography}{1}
\providecommand{\url}[1]{#1}
\csname url@rmstyle\endcsname
\providecommand{\newblock}{\relax}
\providecommand{\bibinfo}[2]{#2}
\providecommand\BIBentrySTDinterwordspacing{\spaceskip=0pt\relax}
\providecommand\BIBentryALTinterwordstretchfactor{4}
\providecommand\BIBentryALTinterwordspacing{\spaceskip=\fontdimen2\font plus
\BIBentryALTinterwordstretchfactor\fontdimen3\font minus
  \fontdimen4\font\relax}
\providecommand\BIBforeignlanguage[2]{{%
\expandafter\ifx\csname l@#1\endcsname\relax
\typeout{** WARNING: IEEEtran.bst: No hyphenation pattern has been}%
\typeout{** loaded for the language `#1'. Using the pattern for}%
\typeout{** the default language instead.}%
\else
\language=\csname l@#1\endcsname
\fi
#2}}

\bibitem{Hu-Courcoubertis1995IEEEComm}
C.~Courcoubetis, G.~Kesidis, A.~Ridder, J.~Walrand, and R.~Weber, ``Admission
  control and routing in {ATM} networks using inferences from measured buffer
  occupancy,'' \emph{IEEE Transactions on Communications}, vol.~43, no. 2/3/4,
  pp. 1778--1784, February/March/April 1995.

\bibitem{Hu-Mao2005ComComm}
G.~Mao, ``A real-time loss performance monitoring scheme,'' \emph{Computer
  Communications}, vol.~28, no.~2, pp. 150--161, February 2005.

\bibitem{Hu-Choe1998}
J.~Choe and N.~Shroff, ``A central-limit-theorem-based approach for analyzing
  queue behavior in high-speed networks,'' \emph{IEEE/ACM Transactions on
  Networking}, vol.~6, no.~5, pp. 659--671, October 1998.

\bibitem{Hu-Kim1998ACM}
H.~S. Kim and N.~B. Shroff, ``Loss probability calculations and asymptotic
  analysis for finite buffer multiplexers,'' \emph{IEEE/ACM Transactions on
  Networking}, vol.~6, no.~4, pp. 411--421, August 1998.

\bibitem{Hu-Knightly1999IEEENetwork}
E.~Knightly and N.~Shroff, ``Admission control for statistical {QoS}: theory
  and practice,'' \emph{IEEE Network}, vol.~13, no.~2, pp. 20--29, March 1999.

\bibitem{Hu-Willinger2002}
W.~Willinger, V.~Paxson, R.~H. Riedi, and M.~S. Taqqu, \emph{Long-range
  dependence: theory and applications}.\hskip 1em plus 0.5em minus 0.4em\relax
  Birkhauser, 2002, ch. Long-range dependence and data network traffic.

\bibitem{Hu-Eun2003TON}
D.~Y. Eun and N.~B. Shroff, ``A measurement-analytic approach for {QoS}
  estimation in a network based on the dominant time scale,'' \emph{IEEE/ACM
  Transactions on Networking}, vol.~11, no.~2, pp. 222--235, April 2003.

\bibitem{Hu-Rice1995Book}
J.~A. Rice, \emph{Mathematical Statistics and Data Analysis}, 2nd~ed.\hskip 1em
  plus 0.5em minus 0.4em\relax Wadsworth Inc., 1995.

\bibitem{Hu-Dan2006Tele}
G.~Dan, V.~Fodor, and G.~Karlsson, ``On the effect of the packet size
  distribution on the packet loss process,'' \emph{Telecommunication Systems},
  vol.~32, no.~1, pp. 31--53, May 2006.

\end{thebibliography}

\end{document}